\documentclass[twocolumn,aps,prl,preprintnumbers]{revtex4}
\usepackage{}
\usepackage{bbm}
\usepackage[latin9]{inputenc}
\usepackage{amsmath}
\usepackage{amssymb}
\usepackage{graphicx}
\usepackage{mathrsfs}
\usepackage{amsfonts}
\usepackage{amsthm}
\usepackage{color}
\usepackage{txfonts}
\usepackage[colorlinks=true,citecolor=blue,linkcolor=blue,urlcolor=blue,anchorcolor=blue]{hyperref}
\hypersetup{colorlinks=true,citecolor=blue,linkcolor=blue,urlcolor=blue}
\providecommand{\U}[1]{\protect\rule{.1in}{.1in}}
\makeatletter
\@ifundefined{textcolor}{}
{
\definecolor{BLACK}{gray}{0}
\definecolor{WHITE}{gray}{1}
\definecolor{RED}{rgb}{1,0,0}
\definecolor{GREEN}{rgb}{0,1,0}
\definecolor{BLUE}{rgb}{0,0,1}
\definecolor{CYAN}{cmyk}{1,0,0,0}
\definecolor{MAGENTA}{cmyk}{0,1,0,0}
\definecolor{YELLOW}{cmyk}{0,0,1,0}
}
\makeatother
\begin{document}
\title{Higher-order exceptional points in all-magnetic structures}
\author{Tianlin Yu}
\author{Huanhuan Yang}
\author{Lingling Song}
\author{Peng Yan}
\email[]{yan@uestc.edu.cn}
\author{Yunshan Cao}
\email[]{yunshan.cao@uestc.edu.cn}
\affiliation{School of Electronic Science and Engineering and State Key Laboratory of Electronic Thin Films and Integrated Devices, University of Electronic Science and Technology of China, Chengdu 610054, China}
\begin{abstract}
Magnetometers with exceptional sensitivity are highly demanded in solving a variety of physical and engineering problems, such as measuring Earth's weak magnetic fields and prospecting mineral deposits and geological structures. It has been shown that the non-Hermitian degeneracy at exceptional points (EPs) can provide a new route for that purpose, because of the nonlinear response to external perturbations. One recent work [H. Yang \emph{et al.,} \href{https://doi.org/10.1103/PhysRevLett.121.197201}{Phys. Rev. Lett. \textbf{121}, 197201 (2018)}] has made the first step to realize the second-order magnonic EP in ferromangetic bilayers respecting the parity-time symmetry. In this paper, we generalize the idea to higher-order cases by considering ferromagnetic trilayers consisting of a gain, a neutral, and a (balanced-)loss layer. We observe both second- and third-order magnonic EPs by tuning the interlayer coupling strength, the external magnetic field, and the gain-loss parameter. We show that the magnetic sensitivity can be enhanced by $3$ orders of magnitude comparing to the conventional magnetic tunneling junction based sensors. Our results pave the way for studying high-order EPs in purely magnetic system and for designing magnetic sensors with ultrahigh sensitivity.
\end{abstract}
\maketitle
\section{Introduction}
The magnetometer, for measuring the intensity of magnetic fields, was firstly created by Carl Friedrich Gauss in 1833 \cite{Gauss1832} and received tremendous progress since then. It has been widely utilized in mineral explorations \cite{Sharma1987,Hato2013}, accelerator physics \cite{Arpaia2018}, archaeology \cite{Fassbinder2016}, mobile phones \cite{Cai2012}, etc. A long-term goal in the community is to pursue magnetometers with ultrahigh sensitivity. Conventional techniques in magnetic sensors encompass many aspect of physics. For example, fluxgate magnetometer works due to the nonlinear character of soft magnetic materials when they are saturated \cite{Primdahl1979,Acuna1974}. Magnetoresistive devices typically are made of thin strips of permalloy whose electrical resistance varies with external magnetic fields \cite{Caruso1997}. Although different magnetometric devices are designed based on different physical mechanisms, they share a general rule that the variation of the order parameter linearly varies with respect to the magnetic field. Presently, ultrahigh-sensitive magnetometers like superconducting quantum interference devices can reach a magnetic sensitivity of $1\text{fT}/\text{Hz}^{1/2}$, but they require an extreme low working temperature and an oversized volume \cite{Gallop2003,Kleiner2004}. Seeking a solid state, small size, room temperature magnetometer with ultrahigh sensitivity is thus one central issue. Recently, it has been demonstrated that the peculiar non-Hermitian degeneracy in magnetic structures \cite{Lee2015,Yang2018,Cao2019,Liu2019,Yuan2020} may provide a promising way to solve the problem.

Hamiltonian obeying the parity-time ($\mathcal{PT}$) symmetry constitutes a special non-Hermitian system, which is invariant under combined parity $\mathcal{P}$ and time-reversal $\mathcal{T}$ operations. It has attracted a lot of attention due to both the fundamental interest in quantum theory \cite{Bender1998, Bender2007, Konotop2016} and the promising application in many fields \cite{Hodaei2017,Feng2014,Hodaei2014}, such as optics \cite{Makris2008,Guo2009,Ruter2010}, tight-binding modeling \cite{Bendix2009,Joglekar2010}, acoustics \cite{Zhu2014,Jing2014}, electronics \cite{Schindler2011,Choi2018}, and very recently in spintronics \cite{Lee2015,Yang2018,Cao2019,Liu2019,Yuan2020}. A $\mathcal{PT}$-symmetric Hamiltonian could exhibit entirely real spectra and a spontaneous symmetry breaking accompanied by a real-to-complex spectra phase transition at the exceptional point (EP) where two or more eigenvalues and their corresponding eigenvectors coalesce simultaneously. In the vicinity of the EP, the eigenfrequency shift follows the $1/N$ power law of the external perturbation, where $N$ is the order of the EP. Such a feature can significantly enhance the sensitivity and has been observed by several experiments \cite{Chen2017,Hodaei2017,Chen2018,Chen20182,Zeng2019}.

$\mathcal{PT}$ symmetry and EP in magnetic systems are receiving growing recent interests. In a simple bilayer structure of two macrospins with balanced gain and loss, the second-order EP (EP2) was observed at a critical Gilbert damping constant \cite{Lee2015}. In Ref. \cite{Zhang2018}, it was proposed to realize the pseudo-Hermiticity in a cavity magnonics system with the third-order EP (EP3). By taking the spin-wave excitation into account, some of the present authors reported a novel ferromagnetic to antiferromagnetic (AFM) phase transition at the EP that depends on magnon's wave vector \cite{Yang2018}. In Ref. \cite{Cao2019}, an exceptional magnetic sensitivity was predicted in the vicinity of the EP3 for $\mathcal{PT}$-symmetric cavity magnon polaritons. However, high-order EPs in purely magnetic/magnonic system is yet to be explored.

In this work, we propose a ferromagnetic trilayer structure consisting of a gain, a neutral, and a loss layer to achieve the EP3. We show that, in the vicinity of the EP3, the separation of eigenfrequencies follow
a power law $\Delta \omega_{\text {EP3}}\propto \epsilon^{1/3}$. Here, the perturbation $\epsilon$ comes from the disturbing magnetic field. We find mode-dependent EPs when the spin-wave excitation is allowed by including the intralayer exchange coupling. A ferromagnetic-to-antiferromagnetic phase transition is observed when the $\mathcal{PT}$ symmetry is broken. Our results suggest a promising way to realize higher-order non-Hermitian degeneracy in a purely magnetic system and to design magnetometer with ultrahigh sensitivities.

The paper is organized as follows: Section \ref{Sec_Macrospins} gives the macrospin model. The condition for observing EP3 is analytically derived. The one-half and one-third power law around EP2 and EP3 are demonstrated, respectively. The effect of noise on the magnetic sensitivity is analyzed as well. In Sec. \ref{Sec_FM_trilayer}, we extend the idea to ferromagnetic trilayers, by allowing spin-wave excitations. Discussion and conclusion are drawn in Sec. \ref{Sec_conclusion}.

\section{Macrospin Model} \label{Sec_Macrospins}
\begin{figure}[h]
  \centering
  \includegraphics[width=0.45\textwidth]{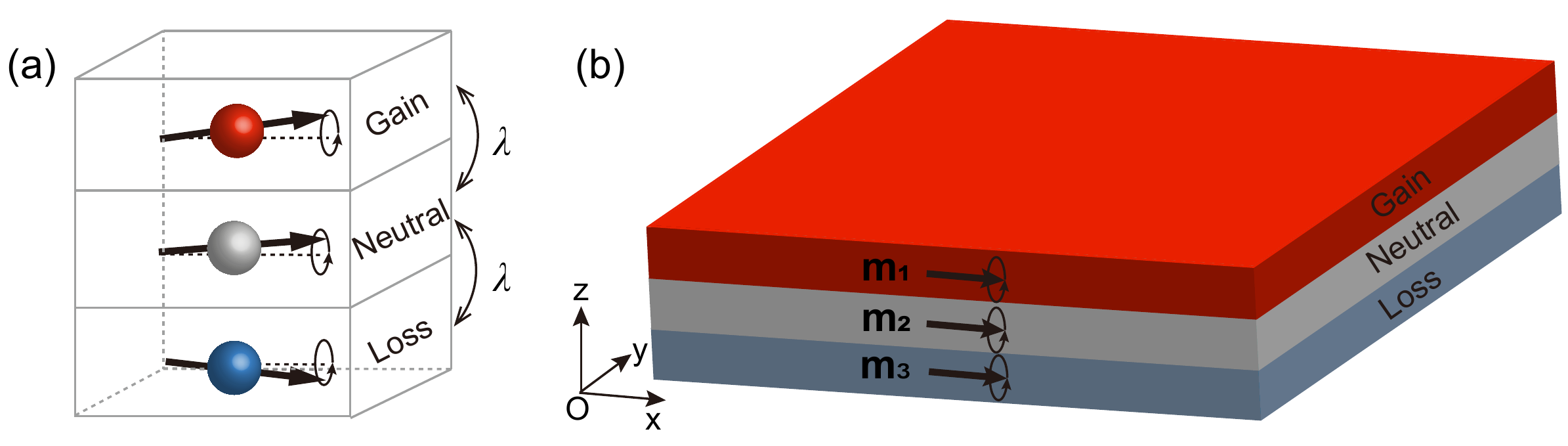}\\
  \caption{(a) Illustration of three exchange-coupled macrospins consisting of a gain (red), neutral (grey), and (balanced-)loss (blue) spin. (b) Schematic plot of a ferromagnetic heterostructure with a gain, neutral, and loss layer, denoted by red, grey, and blue colors, respectively. The magnetizations of all spins are initially along ${\hat x}$-direction.}\label{model}
\end{figure}
We first consider a ternary macrospin structure shown in Fig. \ref{model}(a). The Hamiltonian contains the Zeeman energy, magnetic anisotropy, and exchange coupling:
\begin{equation}
\mathcal{H}=-\sum_n{\bf B}\cdot{\bf M}_{n}-\sum_n\frac{K_n}{2}(m^x_{n})^2
-\lambda\mu_0{\bf M}_{2}\cdot({\bf M}_{1}+{\bf M}_{3}),
\end{equation}
where ${\bf M}_{n}$ (${\bf m}_{n}={\bf M}_{n}/M_n$) is the spin (unit spin) with subscript index $n$ labeling the $n$-th layer ($n=1,2,3$), $M_n$ is the saturated magnetization, ${\bf B}=B\hat{x}$ is the external magnetic field applied on the whole structure, $K_n>0$ is the uniaxial anisotropy, ${\lambda>0}$ is the ferromagnetic exchange-coupling strength between two adjacent layers, and $\mu_0$ is the vacuum permeability. The top and bottom layers are assumed to be the same material but with opposite Gilbert damping parameters, to guarantee the $\mathcal{PT}$ symmetry. The coupled magnetization dynamics is described by the Landau-Lifshitz-Gilbert (LLG) equation \cite{Wang20091,Wang20092}:
\begin{subequations}\label{LLG}
\begin{eqnarray}
\frac{\partial {\bf m}_{1}}{\partial t}&=&-\gamma{\bf m}_{1}\times{\bf B}_\text{eff,1}-\alpha{\bf m}_{1}\times \frac{\partial {\bf m}_{1}}{\partial t},\\
\frac{\partial {\bf m}_{2}}{\partial t}&=&-\gamma{\bf m}_{2}\times{\bf B}_\text{eff,2},\\
\frac{\partial {\bf m}_{3}}{\partial t}&=&-\gamma{\bf m}_{3}\times{\bf B}_\text{eff,3}+\alpha{\bf m}_{3}\times \frac{\partial {\bf m}_{3}}{\partial t},
\end{eqnarray}
\end{subequations}
where $\gamma$ is the gyromagnetic ratio, $\alpha>0$ is the Gilbert constant employed as the balanced gain-loss parameter. The effective magnetic fields read:
\begin{subequations}\label{Heff1}
\begin{eqnarray}
{\bf B}_\text{eff,1}&=&{B}\hat{x}+\frac{K_1}{M_1}{m}_{1}^x\hat{x}+\lambda\mu_0M_2{\bf m}_{2},\\
{\bf B}_\text{eff,2}&=&{B}\hat{x}+\frac{K_2}{M_2}{ m}_{2}^x\hat{x}+\lambda\mu_0M_1({\bf m}_{1}+{\bf m}_{3}),\\
{\bf B}_\text{eff,3}&=&{B}\hat{x}+\frac{K_1}{M_1}{ m}_{3}^x\hat{x}+\lambda\mu_0M_2{\bf m}_{2}.
\end{eqnarray}
\end{subequations}
For small-amplitude spatiotemporal magnetization precession, we assume ${\bf m}_{n}=\hat{x}+m^y_{n}\hat{y}+m^z_{n}\hat{z}$ with $|m^{y,z}_{n}| \ll 1$. By substituting Eqs. (\ref{Heff1}) into Eqs. (\ref{LLG}), and introducing $\psi_{n}=m^y_{n}-im^z_{n}$, we obtain
\begin{subequations}
\begin{eqnarray}
  (i+\alpha)\dot{\psi}_{1}&=&
\omega_1\psi_{1}-\omega_{\lambda2}\psi_{2},\\
i\dot{\psi}_{2}&=&-\omega_{\lambda1}\psi_{1}+\omega_2\psi_{2}-\omega_{\lambda1}\psi_{3},\\
(i-\alpha)\dot{\psi}_{3}&=&-\omega_{\lambda2}\psi_{2}+\omega_1\psi_{3},
\end{eqnarray}
\end{subequations}
where $\omega_1=\gamma(B+K_1/M_1+\lambda\mu_0M_2)$, $\omega_2=\gamma(B+K_2/M_2+2\lambda\mu_0M_1)$, $\omega_{\lambda1}=\gamma\lambda \mu_0 M_1$, and $\omega_{\lambda2}=\gamma\lambda \mu_0 M_2$.
Imposing a harmonic time-dependence $\psi_{n}=\phi_{n}\exp(-i \omega t)$, we have the secular equation:
\begin{equation} \label{eigenvalueEq}
\omega \phi= H \phi,
\end{equation}
with $\phi=(\phi_1,\phi_2,\phi_3)^\text{T}$, and
\begin{equation} \label{eig}
H=\left(
                       \begin{array}{ccc}
                         \frac{\omega_1}{1-i\alpha} & -\frac{\omega_{\lambda2}}{1-i\alpha} & 0 \\
                         -\omega_{\lambda1} & \omega_2 & -\omega_{\lambda1} \\
                         0 & -\frac{\omega_{\lambda2}}{1+i\alpha} & \frac{\omega_1}{1+i\alpha} \\
                       \end{array}
                     \right).
\end{equation}

\subsection{Eigensolutions}
The eigenfrequencies are determined by the zeros of the characteristic polynomial of (\ref{eig}):
\begin{equation} \label{CP}
a\omega^3+b\omega^2+c\omega+d=0,
\end{equation}
with $a=-(1+\alpha^2)<0$, $b=2\omega_1+(1+\alpha^2)\omega_2$, $c=2\omega_{\lambda1}\omega_{\lambda2}-\omega_1^2-2
\omega_1\omega_2$, and $d=\omega_1^2\omega_2-2\omega_1\omega_{\lambda1}\omega_{\lambda2}$. It is known that if and only if $\mathcal{A=B}=0$, the equation has a triple real root, where $\mathcal{A}=b^2-3ac$ and $\mathcal{B}=bc-9ad$. We therefore arrive at the constraint supporting the EP3:
\begin{widetext}
\begin{subequations}\label{AB}
\begin{eqnarray}
&&(2\omega_1+\omega_2+ \alpha^2\omega_2)^2 + 3(1+\alpha^2)(2\omega_{\lambda1}\omega_{\lambda2}-\omega_1^2 - 2\omega_1\omega_2)=0,\\
&&(2\omega_1+\omega_2+\alpha^2\omega_2)(2\omega_{\lambda1}\omega_{\lambda2}-\omega_1^2 - 2\omega_1\omega_2)+
9(1+\alpha^2)(\omega_1^2\omega_2-2\omega_1\omega_{\lambda1}\omega_{\lambda2})=0.
\end{eqnarray}
\end{subequations}
\end{widetext}
To obtain reasonable $\alpha$ and $B$, we note that the difference between $\omega_2$ and $\omega_1$ should be close to $\omega_{\lambda2}$. In the calculations, we thus choose the annealed and deposited Co$_{40}$Fe$_{40}$B$_{20}$ \cite{Burrowes2013,Huo2000} as the top- (bottom-) and the middle-layer materials, with the saturation magnetization $M_1=1.098\times10^6$ A/m and $M_2=1.003\times10^6$ A/m, and the anisotropy constant $K_1=4.36\times 10^5$ J/m$^3$ and $K_2=1.07\times 10^5$ J/m$^3$, respectively.

\begin{figure}
  \centering
  \includegraphics[width=0.48\textwidth]{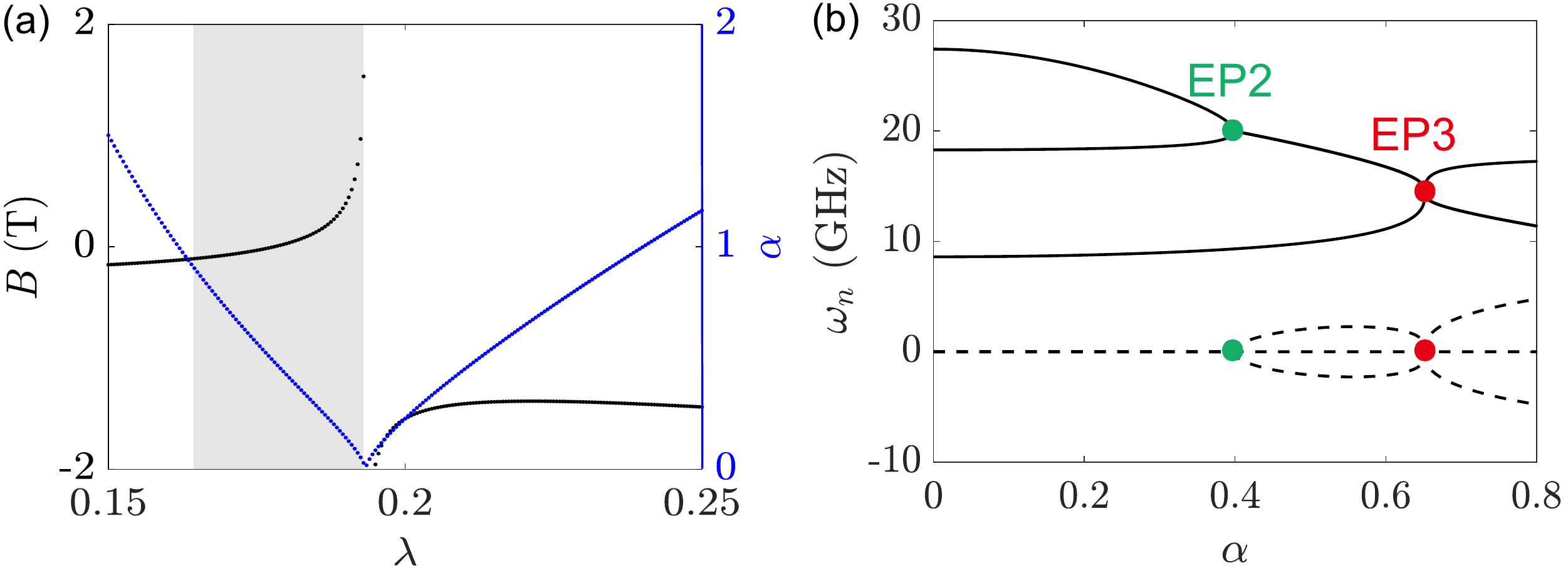}\\
  \caption{(a) Parametric space for EP3. The gray region marks the allowed values of the external magnetic field $B$, the gain-loss parameter $\alpha$, and the interlayer coupling strength $\lambda$. (b) Evolution of eigenvalues as the gain-loss parameter $\alpha$ for $\lambda=0.18$ and $B=29.2$ mT. The solid and dashed curves represent the real and imaginary parts of eigenfrequencies, respectively. }\label{macrospin}
\end{figure}

For each $\lambda$, we numerically calculate the allowed magnetic field $B$ and gain-loss parameter $\alpha$, as shown in Fig. \ref{macrospin}(a) with the black and blue curves, respectively. We note that $B>\text{max}\left\{-K_1/M_1,-K_2/M_2\right\}$ should be satisfied to guarantee a stable ferromagnetic ground state, leading to the reasonable parameters labeled by the gray region in Fig. \ref{macrospin}(a). From Fig. \ref{macrospin}(a), we can see that the critical alpha (magnetic field) decreases (increases) with the increasing of $\lambda$. Figure \ref{macrospin}(b) shows a typical evolution of eigenvalues as the gain-loss parameter $\alpha$ for $\lambda = 0.18$ and $B=29.2$ mT, in which both the EP2 and EP3 emerge, marked by green and red dots, respectively.

Next, we discuss the magnetic sensitivity in the vicinity of EP2 and EP3. The gain-loss parameters $\alpha_{\text{EP2}}=0.399$ and $\alpha_{\text{EP3}}=0.652$ are chosen in the following calculations.

\subsection{Perturbing the top spin}
 Supposing a perturbation $\epsilon$ only on the top macrospin, induced by an external magnetic field $B_{\epsilon}$, i.e., $\epsilon=\gamma B_{\epsilon}/\omega_{\lambda2}$, we modify Eq. (\ref{eigenvalueEq}) to:
\begin{equation} \label{OE}
\Omega \phi= H_{\epsilon} \phi,
\end{equation}
with
\begin{equation}
H_{\epsilon}={\omega_{\lambda2}}\left(
                       \begin{array}{ccc}
                         \frac{\omega_1/\omega_{\lambda2}+\epsilon}{1-i\alpha} & -\frac{1}{1-i\alpha} & 0 \\
                         -\frac{\omega_{\lambda1}}{\omega_{\lambda2}} & \frac{\omega_2}{\omega_{\lambda2}} & -\frac{\omega_{\lambda1}}{\omega_{\lambda2}} \\
                         0 & -\frac{1}{1+i\alpha} & \frac{\omega_1/\omega_{\lambda2}}{1+i\alpha} \\
                       \end{array}
                     \right).
\end{equation}

\begin{figure}
  \centering
  \includegraphics[width=0.48\textwidth]{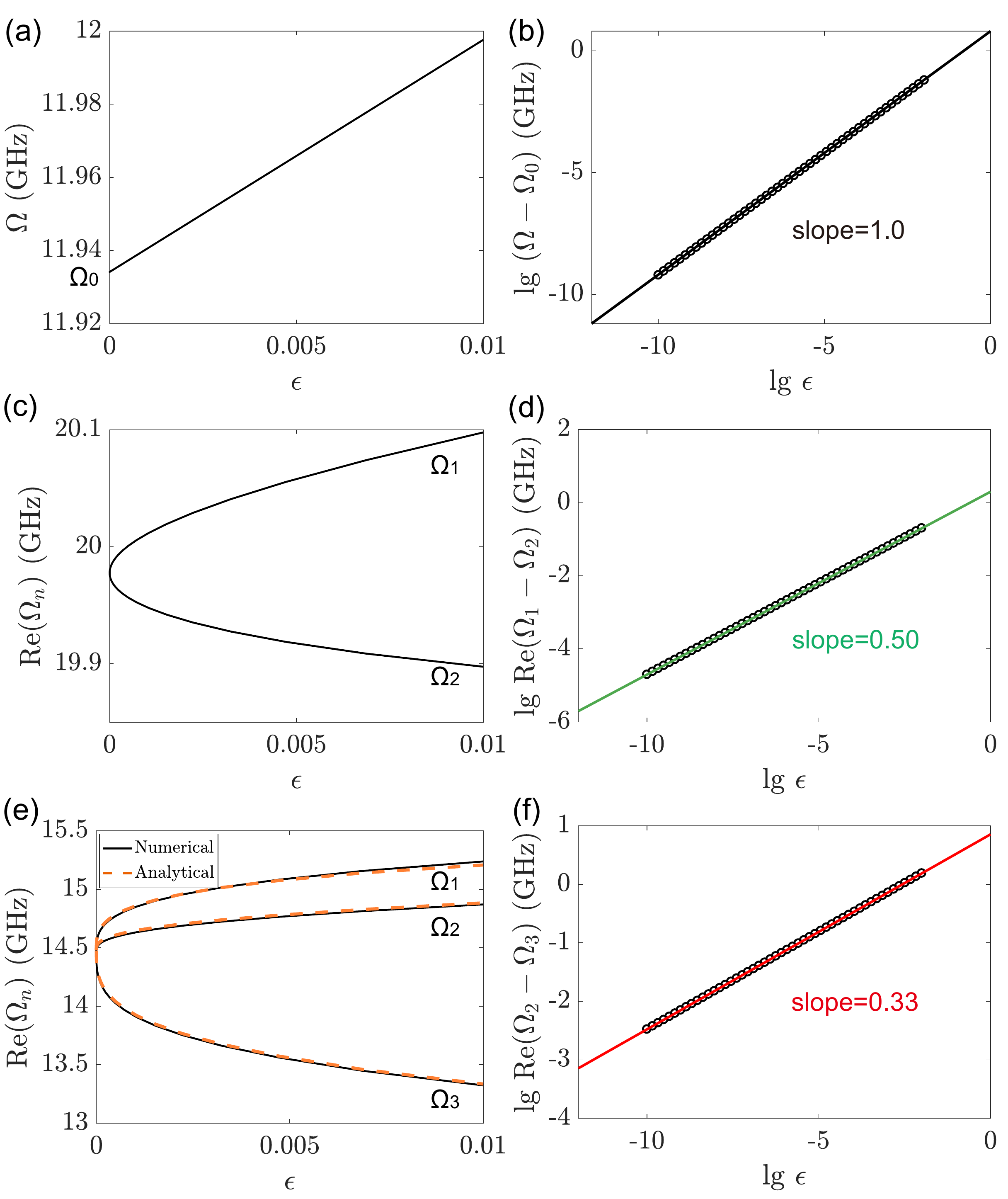}\\
  \caption{(a) The FMR frequency for a single layer ferromagnet as a function of the perturbation $\epsilon$. (b) The frequency shift $\Omega-\Omega_0$ is depicted in logarithmic coordinates, with the slope being 1. (c) The variation of eigenfrequencies near the EP2 as a function of the perturbation. (d) Frequency splitting Re($\Omega_1-\Omega_2$) on a logarithmic scale, with the one-half slope indicating the $\epsilon^{1/2}$ response. (e) The splitting of eigenfrequencies near EP3 v.s. the perturbation. Solid and dashed curves represent numerical and analytical results, respectively. (f) Frequency splitting of Re($\Omega_2-\Omega_3$) on a logarithmic scale, with the slope approximately being $0.33$, suggesting the $\epsilon^{1/3}$ response.}\label{macrospin2}
\end{figure}

To highlight the key role played by the order of the EP, we firstly investigate the effect of the perturbation on a single layer ferromagnet with $\epsilon$ ranging from $10^{-10}$ to $10^{-2}$. We find that the ferromagnetic resonance (FMR) frequency varies linearly with respect to the perturbation plotted in Figs. \ref{macrospin2}(a) and \ref{macrospin2}(b), as naturally expected. Then, we evaluate the variation of eigenvalues with respect to the perturbation near the EP2 and EP3, as depicted in Fig. \ref{macrospin2}(c) and  Fig. \ref{macrospin2}(e), with the mode splitting on a logarithmic scale being plotted in Fig. \ref{macrospin2}(d) and Fig. \ref{macrospin2}(f), respectively. We numerically demonstrate that the separation of frequencies scales as $\epsilon^{1/2}$ and $\epsilon^{1/3}$ for EP2 and EP3, respectively. To have a quantitative comparison, we choose $\epsilon=0.005$ and calculate the frequency difference. We identify 0.03 GHz, 0.14 GHz, and 1.23 GHz shift for the normal FMR, EP2, and EP3 mode, respectively. The sensitivity is thus enhanced by 4.7 and 41 times around EP2 and EP3 with respect to the FMR mode, respectively.

In the following, we analytically derive the frequency splitting near the EP3, by perturbatively solving the characteristic equation of $H_\epsilon$. Based on the Newton-Puiseux series \cite{Barroso2017}, we obtain:
{\begin{equation}\label{spliting}
\frac{{\Omega}_{n}}{{\omega_{\lambda2}}}=c_0+c_{n1}\epsilon^{\frac{1}{3}}
+c_{n2}\epsilon^{\frac{2}{3}}
+c_{n3}\epsilon,\\
\end{equation}
with complex coefficients $c_{ni}$ $(i=1,2,3)$ \cite{coef} and $c_0=2.28$. Solutions (\ref{spliting}) are depicted with dashed orange curves in Fig. \ref{macrospin2}(e), showing a nice agreement with numerical results. The (real part) frequency splitting between $\Omega_1$, $\Omega_2$, and $\Omega_3$ is thus
\begin{equation}\label{solve}
\begin{aligned}
\text{Re}\left(\Omega_1-\Omega_2\right)&=\omega_{\lambda2}\left(0.4\epsilon^{\frac{1}{3}}-0.62\epsilon^{\frac{2}{3}}-0.64\epsilon\right),\\
\text{Re}\left(\Omega_1-\Omega_3\right)&=\omega_{\lambda2}\left(1.53\epsilon^{\frac{1}{3}}-0.61\epsilon^{\frac{2}{3}}-0.64\epsilon\right),\\
\text{Re}\left(\Omega_2-\Omega_3\right)&=\omega_{\lambda2}\left(1.13\epsilon^{\frac{1}{3}}+0.01\epsilon^{\frac{2}{3}}\right),
\end{aligned}
\end{equation}with the leading terms diverging as $\epsilon^{1/3}$, i.e.,
\begin{equation} \label{Delta_O}
\Delta\Omega_{\text {EP3}}=c\omega_{\lambda2} {\epsilon^{1/3}},
\end{equation}
for the separation of $\Omega_2$ and $\Omega_3$ spectral lines with $c=c_{21}-c_{31}$.

Supposing the frequency resolution $|\Delta\Omega_{\text{EP3}}|\approx \kappa_c$, where $\kappa_c$ is bandwidth, we can express the magnetic sensitivity as
\begin{equation} \label{se}
\begin{aligned}
|\delta_B|\approx\frac{\kappa_c}{\gamma C},
\end{aligned}
\end{equation}
where $C\approx \omega_{\lambda2}^2/\kappa^2_c$ is the cooperativity.
Using the following parameters: the damping constant 0.012, the FMR frequency 3.1 GHz, $\kappa_c\approx0.01$ GHz, and $\omega_{\lambda2}=6.35$ GHz, we estimate the sensitivity as $10^{-13}$ T$/\text{Hz}^{1/2}$, which is 3 orders of magnitude higher than the conventional magnetic sensor based on magnetic tunneling junction \cite{Cardoso2014}.

\subsection{Perturbing the whole structure}
\begin{figure}
  \centering
  \includegraphics[width=0.48\textwidth]{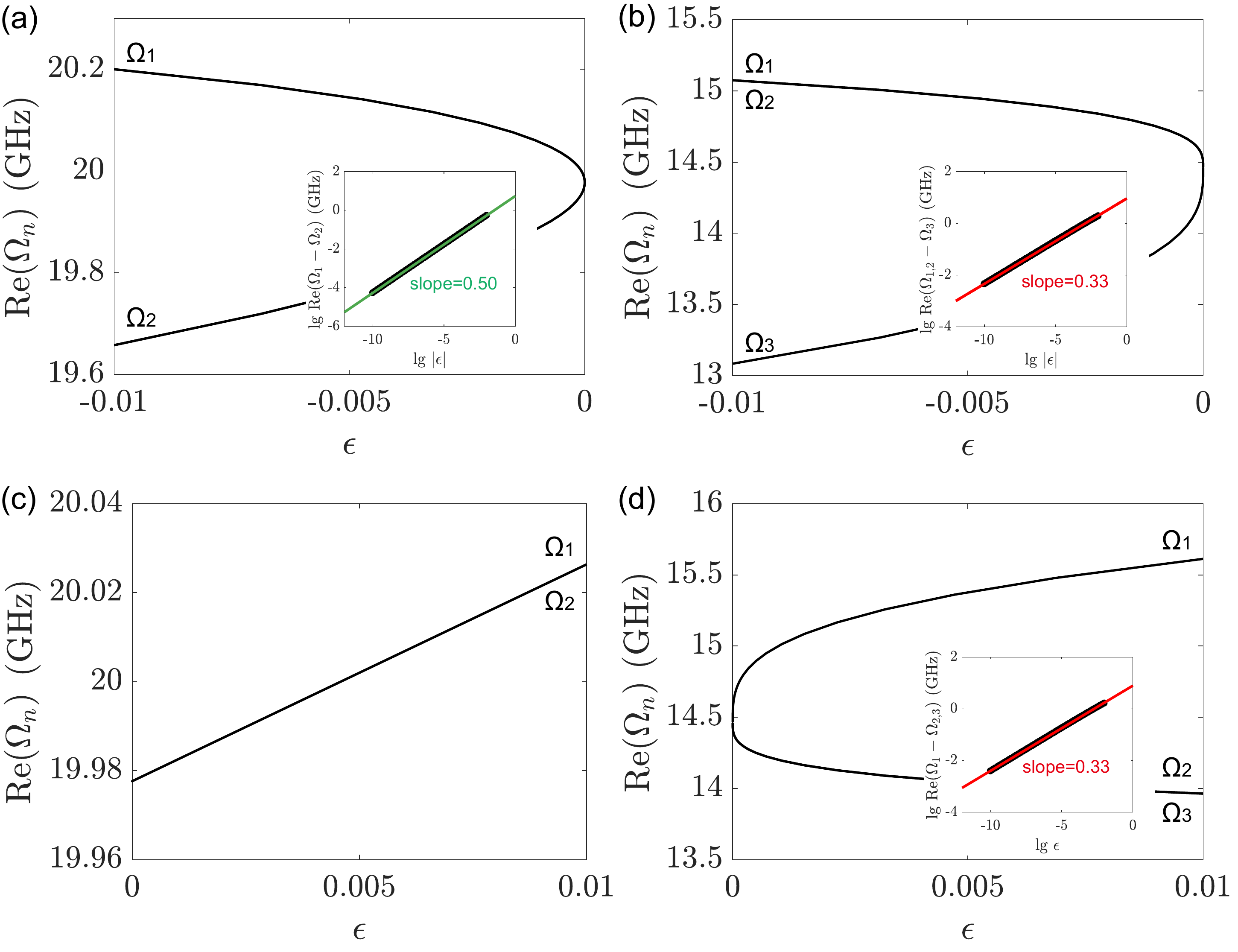}\\
  \caption{Evolution of the eigenfrequencies as a function of the perturbation near the EP2 (a) and EP3 (b) for $\epsilon<0$. Inset: frequency splitting Re($\Omega_1-\Omega_2$) and Re($\Omega_{1,2}-\Omega_3$) on a logarithmic scale, with the slopes approximately being $0.5$ and $0.33$, respectively. Evolution of the eigenfrequencies as a function of the perturbation near the EP2 (c) and EP3 (d) for $\epsilon>0$. Inset plots the frequency splitting on a logarithmic scale.}\label{macrospinA}
\end{figure}

In Sec. \ref{Sec_Macrospins}B, we have considered perturbations only on the top spin. Because of the nonlocal nature of the magnetic field, it may affect the whole macrospin system. For such cases, we re-write the matrix $H_{\epsilon}$:
\begin{equation}
H'_{\epsilon}=\omega_{\lambda2}\left(
                       \begin{array}{ccc}
                        \frac{\omega_1/\omega_{\lambda2}+\epsilon}{1-i\alpha}& -\frac{1}{1-i\alpha} & 0 \\
                         -\frac{\omega_{\lambda1}}{\omega_{\lambda2}} & \frac{\omega_2}{\omega_{\lambda2}}+\epsilon & -\frac{\omega_{\lambda1}}{\omega_{\lambda2}} \\
                         0 & -\frac{1}{1+i\alpha}  & \frac{\omega_1/\omega_{\lambda2}+\epsilon}{1+i\alpha} \\
                       \end{array}
                     \right).
\end{equation}
As shown in Fig. \ref{macrospinA}(a), the eigenfrequency near the EP2 splits into two branches for ${\epsilon}<0$, with the inset displaying the one-half power law behaviour. The frequency near the EP3 splits to two branches as well including two degenerate modes. The separation of two frequencies follows the one-third power law, as plotted in Fig. \ref{macrospinA}(b). For ${\epsilon}>0$, the solutions contain a real root and a pair of complex conjugated roots. The perturbation pushes the spectrum into the exact $\mathcal{PT}$ phase region and thus can't remove the degeneracy of EP2, as depicted in Fig. \ref{macrospinA}(c). Figure \ref{macrospinA}(d) shows the frequency splitting in the vicinity of EP3, which is similar to that shown in Fig. \ref{macrospinA}(b).

When the whole trilayer structure is perturbed for ${\epsilon}>0$, we find the sensitivity approximately to be $10^{-13}$ T Hz$^{-1/2}$, which is the same order of magnitude as the case studied in Sec. \ref{Sec_Macrospins}B.

\subsection{The effect from statistical noise}
\begin{figure}
  \centering
  \includegraphics[width=0.45\textwidth]{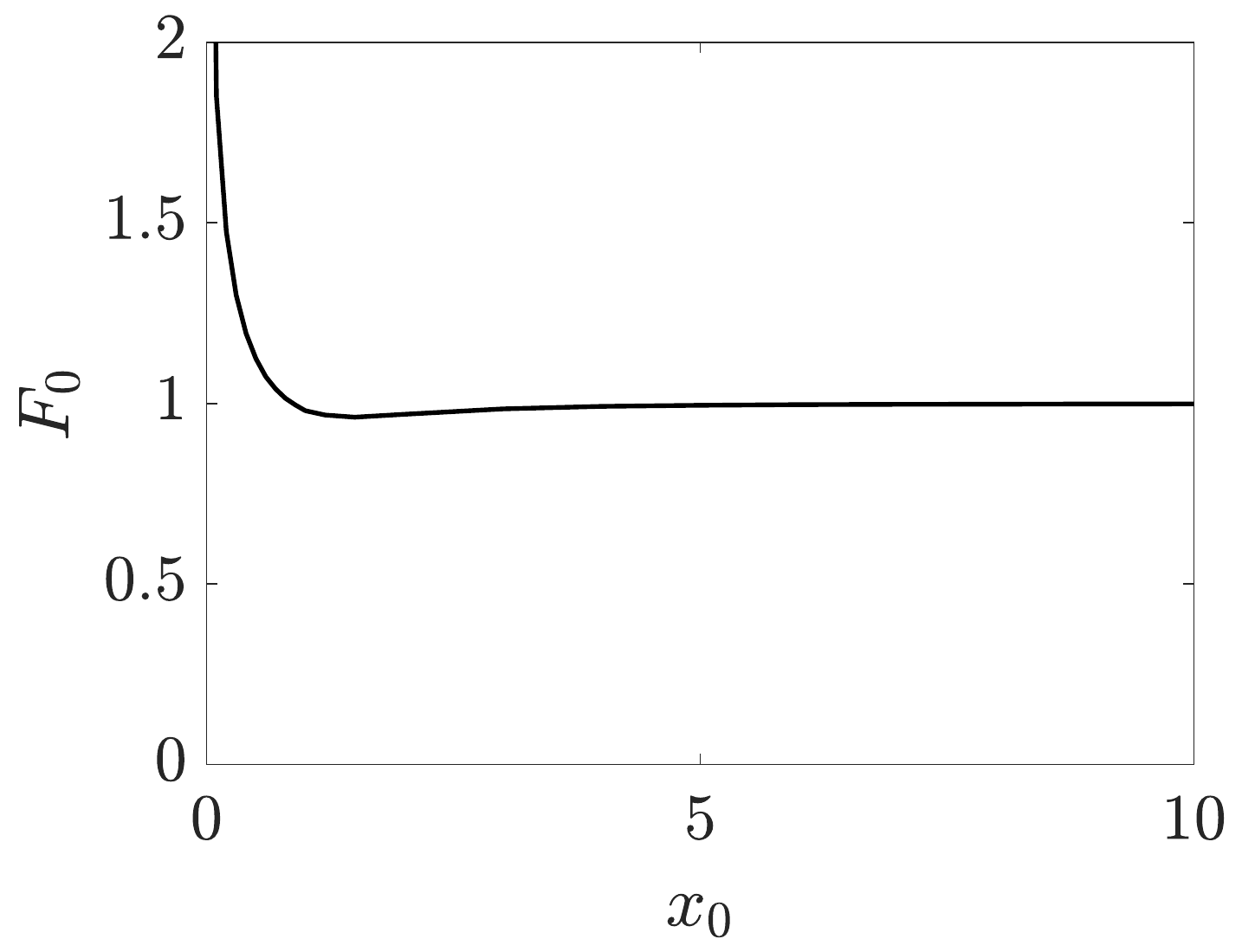}\\
  \caption{Sensitivity-diminution factor $F_0$ as a function of $x_0$.}\label{noise}
\end{figure}
Noise is inevitable in magnetic systems, which may be caused by material imperfections or fluctuating environments. Following the method in Ref. \cite{Cao2019,Mortensen2018}, we consider a Gaussian distribution of the perturbation $\epsilon$:
\begin{equation}
P(\epsilon-\epsilon_0)=\frac{1}{\sqrt{2\pi}\sigma}\exp\left[-\frac{1}{2}\left(\frac{\epsilon-\epsilon_0}
{\sigma}\right)^2\right],
\end{equation}
with the signal $\epsilon_0$ to be detected and the noise level $\sigma$. The ensemble-average sensitivity can be obtained by:
\begin{equation}
\begin{aligned}
\left<\Delta\Omega_{\text {EP3}}\right>&=\int^{+\infty}_{-\infty}c\omega_{\lambda2}\sqrt[3]\epsilon P(\epsilon-\epsilon_0)d\epsilon \\
&=\frac{c\omega_{\lambda2}\sigma^{1/3}}{\sqrt{2\pi}}\int^{+\infty}_{-\infty}|x+x_0|^{1/3}
e^{-\frac{1}{2}x^2}dx,
\end{aligned}
\end{equation}
with $x=(\epsilon-\epsilon_0)/\sigma$ and $x_0=\epsilon_0/\sigma$. In the small and large siginal/noise ratio limit, we obtain
\begin{equation}
 \left<\Delta\Omega_{\text {EP3}}\right>=\left\{
\begin{aligned}
&\frac{2^{1/6}c\omega_{\lambda2}\sigma^{1/3}}{\sqrt{\pi}}\Gamma{\left(\frac{2}{3}\right)},~~~x_0\ll1\\
&c\omega_{\lambda2}\epsilon_0^{1/3},~~~x_0\gg1
 \end{aligned}
  \right.
 \end{equation}

For a large signal/noise ratio, $\left<\Delta\Omega_{\text {EP3}}\right>$ recovers Eq. (\ref{Delta_O}). By defining the sensitivity-diminution factor $F_0=c^{-1}\omega_{\lambda2}^{-1}\epsilon_0^{-1/3}\left<\Delta\Omega_{\text {EP3}}\right>$, we can evaluate the influence of noise on the sensitivity, which is plotted in Fig. \ref{noise}. It shows that the sensor performs well when $x_0>1$.

\section{Trilayer ferromagnetic films} \label{Sec_FM_trilayer}
In this section, we extend the macrospin model to trilayer ferromagnets which include both intralayer and interlayer exchange couplings, as shown in Fig. \ref{model}(b). The Hamiltonian of the system is then given by:
\begin{equation} \label{H2}
\begin{aligned}
\mathcal{H}=&-\sum_n\sum_{<i,j>}J_n{\bf m}_{n,i}\cdot{\bf m}_{n,j}-
\sum_n\sum_{i}{\bf B}_{n,i}\cdot{\bf M}_{n,i}\\
&-\sum_n\sum_{i}\frac{K_n}{2}(m^x_{n,i})^2
-\lambda\mu_0\sum_i{\bf M}_{2,i}\cdot({\bf M}_{1,i}+{\bf M}_{3,i}),
\end{aligned}
\end{equation}
\begin{figure}
  \centering
  \includegraphics[width=0.48\textwidth]{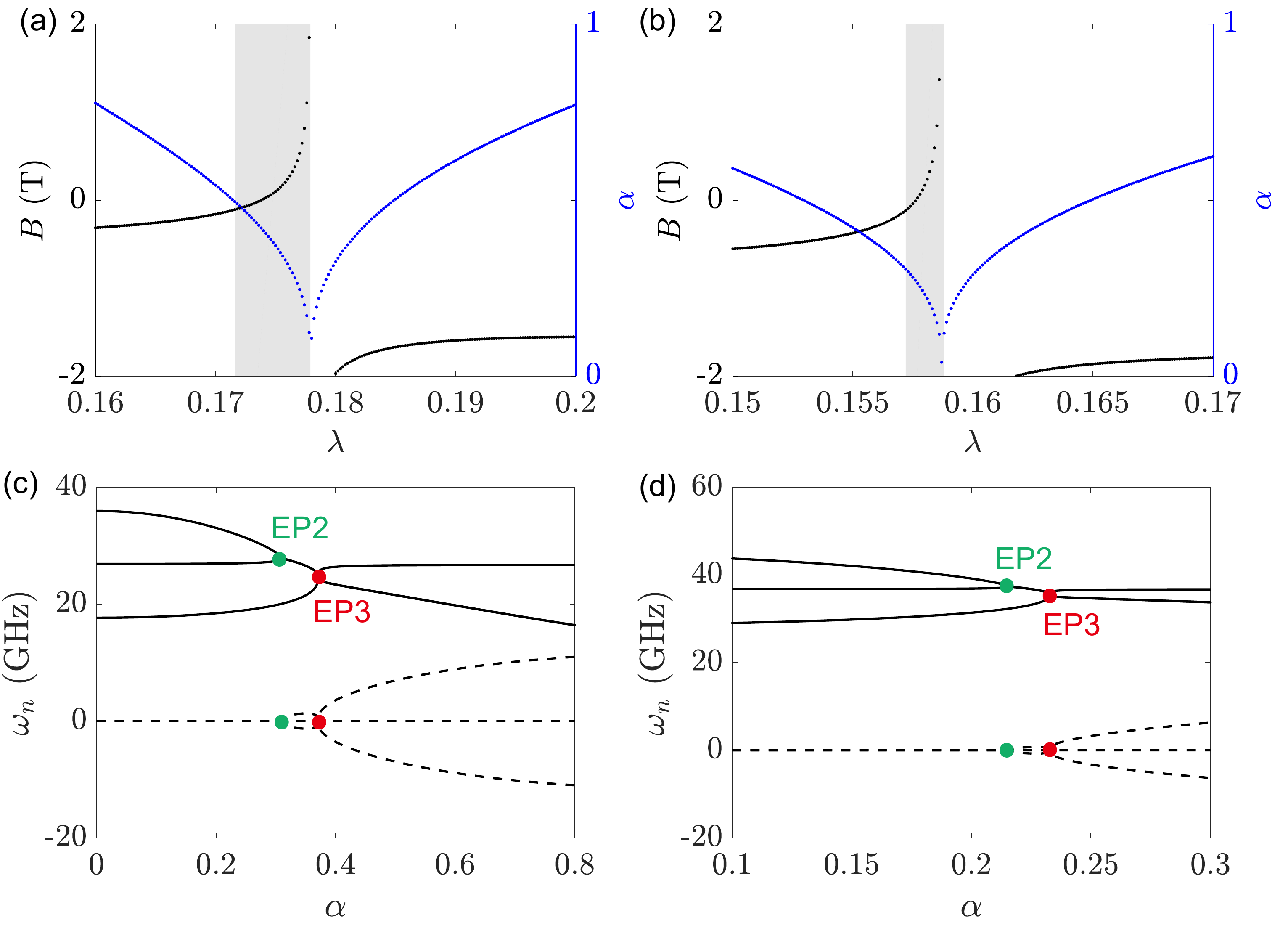}\\
  \caption{The external magnetic field and gain-loss parameter dependence on the interlayer coupling strength $\lambda$ at EP3 for $(k_x,k_y)=(\frac{\pi}{30a}, 0)$ (a) and $(\frac{\pi}{20a}, 0)$ (b). The gray region marks the parametric space allowing the EP3. (c) Evolution of eigenvalues with respect to the gain-loss parameter $\alpha$ for $\lambda=0.175$ and $B=99 $ mT at $(k_x,k_y)=(\frac{\pi}{30a}, 0)$. (d) The real and imaginary parts of the eigenvalues as a function of the gain-loss parameter $\alpha$ for $\lambda=0.158$ and $B=170 $ mT at $(k_x,k_y)=(\frac{\pi}{20a}, 0)$. }\label{trilayers}
\end{figure}
where ${\bf M}_{n,i}$ (${\bf{m}}_{n,i}={\bf M}_{n,i}/M_{n,i}$) is the spin (unit spin) at the $i$-th site in the $n$-th layer ($n=1,2,3$) with the saturation magnetization $M_{n,i}$, $J_n>0$ is the intralayer exchange coupling constant, $\langle i,j\rangle$ sums over all nearest-neighbor sites in the same layer, and ${\bf B}_{n,i}=B_{n,i}\hat{x}$ is the external magnetic field at the $i$-th site in the $n$-th layer. In the calculations, we adopt the same material parameters as the macrospin model and consider the intralayer exchange coupling constant $J_{1,2,3}=J=2.44\times 10^{7}$ J/m$^3$. A homogeneous magnetic field is assumed to be applied over the whole system, i.e., $B_{1,i}=B_{2,i}=B_{3,i}=B$.

The magnetization dynamics is described by the LLG equation (\ref{LLG}) but with the following effective fields:
\begin{equation}\label{Heff} \small
\begin{aligned}
{\bf B}_\text{eff,1,i}&=\frac{J}{M_1}\sum_{<i,j>}{\bf m}_{1,j}+B \hat{x}+\frac{K_1}{M_1}{m}_{1}^x\hat{x}+\lambda\mu_0M_2{\bf m}_{2,i},\\
{\bf B}_\text{eff,2,i}&=\frac{J}{M_2}\sum_{<i,j>}{\bf m}_{2,j}+B \hat{x}+\frac{K_2}{M_2}{m}_{2}^x\hat{x}+\lambda\mu_0M_1({\bf m}_{1,i}+{\bf m}_{3,i}),\\
{\bf B}_\text{eff,3,i}&=\frac{J}{M_1}\sum_{<i,j>}{\bf m}_{3,j}+B \hat{x}+\frac{K_1}{M_1}{m}_{3}^x\hat{x}+\lambda\mu_0M_2{\bf m}_{2,i},
\end{aligned}
\end{equation}
where $J\sum_{<i,j>}{\bf m}_{n,j}$ represents $J[{\bf m}_{n,(i_x-1)a,i_ya}+{\bf m}_{n,(i_x+1)a,i_ya}+{\bf m}_{n,i_xa,(i_y-1)a}+{\bf m}_{n,i_xa,(i_y+1)a}]$ with
$(i_xa, i_ya)$ being the coordinate of the $i$-th unit spin vector, $i_{x(y)}$ is
an integer, and $a$ is the lattice constant.

Considering a small-angle dynamics, we set ${\bf m}_{n,i}=\hat{x}+m^y_{n,i}\hat{y}+m^z_{n,i}\hat{z}$
 with $|m^{y,z}_{n,i}| \ll 1$. Substituting the effective field into Eqs. (\ref{LLG}) and imposing the complex scalar-fields $\psi_{n,i}=m^y_{n,i}-im^z_{n,i}$, we obtain:
\begin{widetext}
\begin{equation}
\begin{aligned}
i\dot{\psi}_{1,i}&= \frac{\gamma J}{M_1}\left(4\psi_{1,i}-\sum_{<i,j>}\psi_{1,j}\right)
+\omega_{\lambda2}\left(\psi_{1,i}-\psi_{2,i}\right)
+\gamma \left(B+\frac{K_1}{M_1}\right)\psi_{1,i}-\alpha\dot{\psi}_{1,i},\\
i\dot{\psi}_{2,i}&= \frac{\gamma J}{M_2}\left(4\psi_{2,i}-\sum_{<i,j>}\psi_{2,j}\right)
+\omega_{\lambda1}\left(2\psi_{2,i}-\psi_{1,i}-\psi_{3,i}\right)+\gamma \left(B+\frac{K_2}{M_2}\right)\psi_{2,i},\\
i\dot{\psi}_{3,i}&=\frac{\gamma J}{M_1}\left(4\psi_{3,i}-\sum_{<i,j>}\psi_{3,j}\right)
+\omega_{\lambda2}\left(\psi_{3,i}-\psi_{2,i}\right)+\gamma \left(B+\frac{K_1}{M_1}\right)\psi_{3,i}+\alpha\dot{\psi}_{3,i},
\end{aligned}
\end{equation}
\end{widetext}
with the abbreviation $\sum_{<i,j>}{\psi}_{n,j} ={\psi}_{n,(i_x-1)a,i_ya}+{ \psi}_{n,(i_x+1)a,i_ya}+{\psi}_{n,i_xa,(i_y-1)a}+{\psi}_{n,i_xa,(i_y+1)a}$.
\begin{figure}
  \centering
  \includegraphics[width=0.48\textwidth]{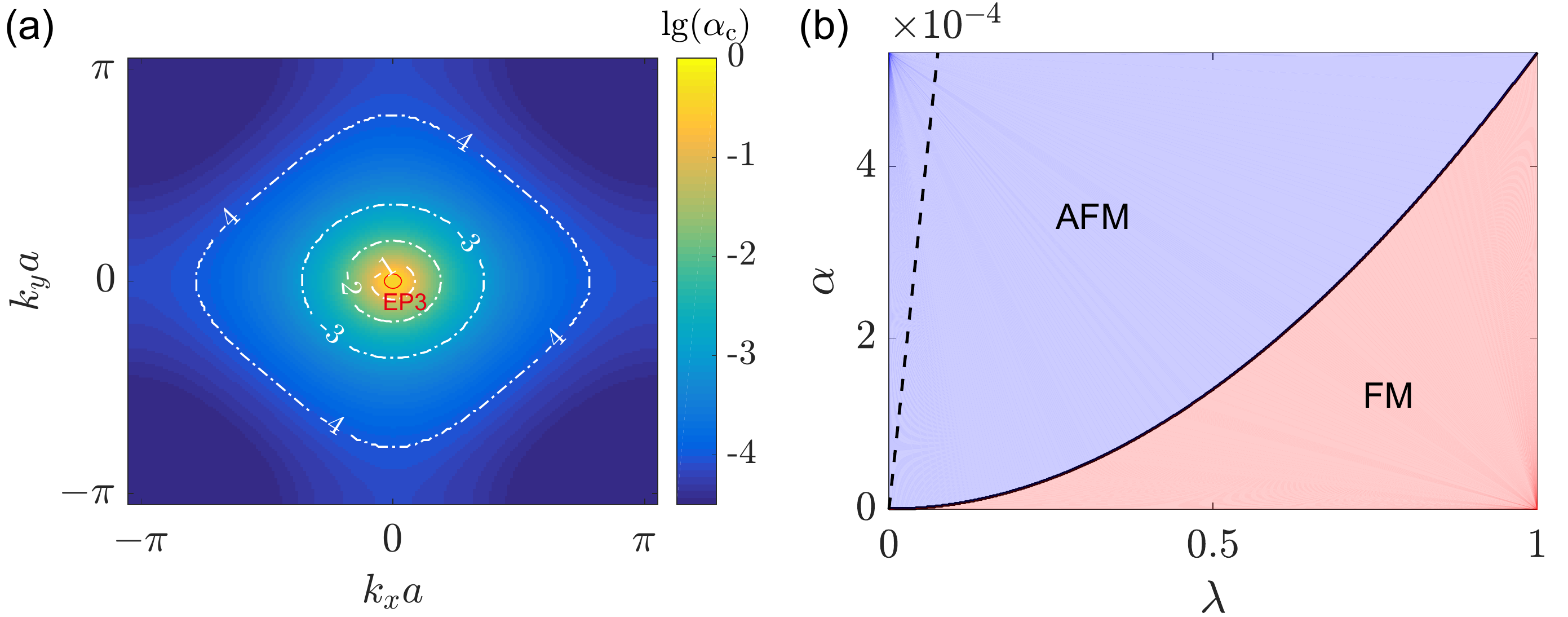}\\
  \caption{(a) Contour plot of the critical gain-loss parameters dependence on spin-wave modes {\bf k}. The parameters are identical to the ones in Fig. \ref{trilayers}(c). (b) FM-AFM phase diagram of the $\mathcal{PT}$-symmetric trilayer and bilayer in $\alpha-\lambda$ plane. The solid and dashed curves represent the phase boundary in the two cases.}\label{alpha_c}
\end{figure}

Expanding the spatiotemporal magnetization in terms of plan waves $\psi_{n,i}=\phi_{n,i}\exp(i{\bf k}\cdot{\bf r}-i \omega t)$, we have:
\begin{equation} \label{feq2}
\omega \phi_i= H_i \phi_i,
\end{equation}
with
\begin{equation}
H_i=\left(
                       \begin{array}{ccc}
                         \frac{\omega'_1}{1-i\alpha} & -\frac{\omega_{\lambda2}}{1-i\alpha} & 0 \\
                         -\omega_{\lambda1} & \omega'_2 & -\omega_{\lambda1} \\
                         0 & -\frac{\omega_{\lambda2}}{1+i\alpha} & \frac{\omega'_1}{1+i\alpha} \\
                       \end{array}
                     \right),
\end{equation} and $\phi_i=(\phi_{1,i},\phi_{2,i},\phi_{3,i})^\text{T}$, where $\omega'_1=\tilde{\omega}_{1}(k_x,k_y)+\gamma(B+K_1/M_1+\lambda\mu_0M_2)$ and $\omega'_2=\tilde{\omega}_{2}(k_x,k_y)+\gamma(B+K_2/M_2+2\lambda\mu_0M_1)$ with $\tilde{\omega}_{n}(k_x,k_y)=2\gamma J/M_n\left[2-\cos(k_xa)-\cos(k_ya)\right]$.

It is straightforward to see that, for $k_x=k_y=0$, Eq. (\ref{feq2}) is reduced to Eq. (\ref{eigenvalueEq}). We aim to search for all EPs in ferromagnetic trilayers. Following Ref. \cite{Yang2018}, we know that the emergence of EP3 depends on magnon's wave vector $\mathbf{k}=(k_x,k_y)$. As two examples, we set $\mathbf{k}=(\frac{\pi}{30a}, 0)$ and $(\frac{\pi}{20a}, 0)$ without loss of generality, to illustrate the condition supporting the EP3, which are depicted in Fig. \ref{trilayers} (a) and Fig. \ref{trilayers}(b), respectively. We then explicitly demonstrate the emergence of EP3 in Fig. \ref{trilayers}(c) and Fig. \ref{trilayers}(d). We observe that the EP2 appears for all spin-wave modes. At a given $(k_x,k_y)$, there exists a critical gain-loss parameter $\alpha_c$, beyond which the exact $\mathcal{PT}$ symmetry is broken. We plot the distribution of the critical gain-loss parameter over the entire Brillouin zone in Fig. \ref{alpha_c}(a). The red circle marks the critical $\alpha$ for the emergence of EP3. In comparison to previous work \cite{Yang2018}, we did not note a special region where the $\mathcal{PT}$ symmetry is never broken. This is due to the fact that the chiral spin-spin coupling, i.e., Dzyaloshinskii-Moriya interaction, is absent in the present model.

As first predicted in Ref. \cite{Yang2018}, for a $\mathcal{PT}$-symmetry ferromagnetic bilayer, antiferromagnetism could emerge in the $\mathcal{PT}$ broken phase. As to the ferromagnetic trilayer, it can exhibit a FM-AFM phase transition as well. In Fig. \ref{alpha_c}(a), we find that the minimum of $\alpha_c$({\bf k}) appears at the boundary of the Brillouin zone. We calculate the corresponding critical gain-loss parameter at ${\bf k}=(\pm \frac{\pi}{a}, \pm \frac{\pi}{a})$ for different $\lambda$,
\begin{equation}
\alpha_c=\sqrt{X(\textbf{k})-1}\left|_{{\bf k}=(\pm \frac{\pi}{a}, \pm \frac{\pi}{a})},\right.
\end{equation}
where
\begin{widetext}
\begin{equation}
X=\frac{1}{12\omega'^3_2\left(\omega'^2_1\omega'_2-2\omega'_1\omega_{\lambda1}\omega_{\lambda2}\right)}\left[c_1+\left(c_3-\sqrt{c_3^2-c_2^3}\right)^{1/3}+\left(c_3+\sqrt{c_3^2-c_2^3}\right)^{1/3}\right],\\
\end{equation}
with
\begin{equation}
\begin{split}
& c_1=-27\beta_1^2-6\beta_1\omega'_2\left(3\beta_2+4\omega'_1\omega'_2\right)+\beta_2^2\omega'^2_2,\\
& c2=\left[27\beta_1^2+6\beta_1\omega'_2\left(3\beta_2+4\omega'_1\omega'_2\right)-\beta_2^2\omega'^2_2\right] ^2-48\beta_1\omega'^3_{2}(9\beta_1\beta_2\omega'_1+12\beta_1\omega'^2_1\omega'_2-\beta_2^3-\beta_2^2\omega'_1\omega'_2),\\
& c3=-\left[27\beta_1^2+6\beta_1\omega'_2\left(3\beta_2+4\omega'_1\omega'_2\right)-\beta_2^2\omega'^2_2\right] ^3+72\beta_1\omega'^3_{2}\left[27\beta_1^2+6\beta_1\omega'_2\left(3\beta_2+4\omega'_1\omega'_2\right)-\beta_2^2\omega'^2_2\right]\\
&\ \ \ \ \ \ \ \ \times\left(9\beta_1\beta_2\omega'_1+12\beta_1\omega'^2_1\omega'_2-\beta_2^3-\beta_2^2\omega'_1\omega'_2\right)-864\beta_1^2\omega'^2_1\omega'^6_{2}\left(8\beta_1\omega'_1-\beta_2^2\right),\\
&\beta_1=\omega'^2_1\omega'_2-2\omega'_1\omega_{\lambda1}\omega_{\lambda2},\\
&\beta_2=2\omega_{\lambda1}\omega_{\lambda2}-\omega'^2_1-2
\omega'_1\omega'_2,
\end{split}
\end{equation}
\end{widetext}
as plotted by the solid black curve in Fig. \ref{alpha_c}(b), in which the blue and red regions represent the AFM and FM phases, respectively. The phase boundary for $\mathcal{PT}$-symmetric bilayer is
\begin{equation}
\alpha_c=\frac{\lambda\mu_0 M_1}{\sqrt{\frac{8J}{M_1}+B+\frac{K_1}{M_1}}\sqrt{\frac{8J}{M_1}+B+\frac{K_1}{M_1}+2\lambda\mu_0 M_1}}
\end{equation}
marked by the dashed line in Fig. \ref{alpha_c}(b), as a comparison.

\section{Discussion and Conclusion}\label{Sec_conclusion}
Negative damping (gain) is the key to realize our proposal. In previous work \cite{Lee2015,Yang2018}, it has been suggested that the spin transfer torque, the parametric driving, the ferromagnetic$|$ferroelectric heterostructure \cite{Jia2015}, and the interaction between magnetic system and environment \cite{Ando2008,Duan2014,Li2014} are possible mechanisms to achieve the magnetic gain. Slavin \emph{et al.} analytically demonstrated that the main effect of the spin-polarized current in a ``free" magnetic layer is a negative damping \cite{Slavin2005}. In Ref. \cite{Apalkov2005}, the Slonczewski form of the spin torque is treated as a negative damping too.

To achieve the EP3, FM coupling between two adjacent layers should fall into the allowed parametric space, which can be realized by tuning the thickness of the nonmagnetic spacer between them \cite{Frackowiak2018,Vincent2002}. A single-mode spin wave can be excited via the Brillouin light scattering technique \cite{Park2002,Dem2001}, which is essential to observe the mode-dependent EP3.

In summary, we have theoretically investigated the dynamics of $\mathcal{PT}$-symmetric ternary macrospin structure and ferromagnetic trilayer. We observed both EP2 and EP3 under proper materials parameters. We demonstrated the one-half and one-third power law response to external perturbations in the vicinity of EP2 and EP3, respectively. Outstanding magnetic sensitivities were identified in the vicinity of EP3. Our results open the door for observing higher-order EPs in all-magnetic structures and for designing ultrahigh-sensitive magnetometers.

\section{acknowledgement}
This work was supported by the National Natural Science
Foundation of China (Grants No. 11704060 and No. 11604041).

\end{document}